
\input harvmac
\input epsf
\def\NP#1{Nucl.\ Phys. {\bf B{#1}}}
\def\PL#1{Phys.\ Lett. {\bf B{#1}}}
\def\PREP#1{Phys.\ Rep. {\bf {#1}}}
\def\MPL#1{Mod. Phys. Lett. {\bf A{#1}}}
\def\PRD#1{Phys.\ Rev. {\bf D{#1}}}
\Title{\vbox{\baselineskip12pt\hbox{}\hbox{September, 1993}}}
{\vbox{\centerline{Phase diagrams of a Higgs-Yukawa system with}
       \vskip3pt
       \centerline{chirally invariant lattice fermion actions}}}

\centerline{Sergei V. Zenkin \footnote{$^*$}{Permanent address: Institute for
Nuclear Research of the Russian Academy of Sciences, Moscow 117312, Russia..E-mail addresses: zenkin@jpnyitp.bitnet, h79104a@kyu-cc.cc.kyushu-u.ac.jp}}
\bigskip \centerline{Department of Physics, Kyushu University 33}
\centerline{Fukuoka 812, Japan}
\vskip 1cm

We study the phase structure of a $Z_2$ lattice Higgs-Yukawa system with
different forms of chirally invariant lattice fermion actions: naive, two
types
of non-local actions, and a mirror fermion action. The calculations are
performed in the variational mean field approximation with contribution of
fermionic determinant being calculated in a ladder approximation. We
demonstrate
that the phase structure of the system crucially depends on the form of the
lattice fermion action, despite the fact that the Higgs field couples to
fermions in the same local way. A possibility of defining chiral gauge
theories
based on mirror fermion action is briefly discussed.
\Date{}

\newsec{Introduction}

The lattice fermion actions which possess global chiral
symmetry is of particular interest for nonperturbative formulation of gauge
theories and the Higgs phenomenon (for resent reviews see \ref\PM{D. N.
Petcher, Proc. of the XXVI Int. Conf. on High Energy Physics (Dallas, August
1992),
Vol. II, ed. J. R. Stanford (AIP No. 272) p. 1529; Nucl. Phys. B (Proc.
Suppl.) 30
(1993) 50;
I. Montvay, Nucl. Phys. B (Proc. Suppl.) 26 (1992) 57}). In this paper we
study
such actions
in a more simple aspect, namely, we investigate how the phase structure of a
Higgs-Yukawa
system depends on their form. As a first step we consider phase diagrams of a
$Z_2$ model. For this sake we use the variational mean field approximation
with
contributions of fermionic determinant  being summed up in a ladder
approximation \ref\Z{S. V. Zenkin, ``On a mean field approximation for
Higgs-Yukawa systems", preprint KYUSHU-HET-7 (1993), hep-lat/9309001}. This
enables us to infer phase diagrams of the system in fact for any values of
the Higgs-Yukawa coupling.

By virtue of the Nielsen-Ninomiya theorem \ref\NN{H. B. Nielsen, M. Ninomiya,
\NP{185} (1981) 20; \NP{193} (1981) 173}, chirally invariant lattice fermion
actions must be either non-Hermitean, or non-local, otherwise they involve an
equal number of the left-handed and right-handed Weyl fermions. As we use a
mean
field method, we cannot take non-Hermitean actions, being able to consider
non-local and mirror fermion actions. Besides, we restrict ourselves by only
those actions which are well-defined on a finite lattice. As a benchmark we
reproduce also a result of ref. [2] for naive fermion action.

The paper is organized as follows. The system is defined in Sect. 2. For a
reader convenience we give the basic points of the method and
approximations of ref. [2] in Sect. 3. In Sect. 4 we apply the formulae of
Sect. 3 to the system with naive, SLAC-like, a maximally non-local, and a
mirror fermion actions. Sect. 5 is a conclusion.

\newsec{The system}

The system is defined on a hypercubic $D$-dimensional ($D$ is even) lattice
$\Lambda$ with sites numbered by $n = (n_1, ..., n_D)$, $-N/2+1 \leq n_{\mu}
\leq N/2$ ($N$ is even) and with lattice spacing $a = 1$; $\hat{\mu}$ is the
unit vector along a lattice link in the positive $\mu$-direction. Dynamical
variables of the model are the fermion $2^{D/2}$-component fields $\psi_n$,
$\psibar_n$, and scalar field $\phi_n \in Z_2$ (i.e. $\phi_n = \pm 1$). We
imply
antiperiodic boundary conditions for the fermion and periodic for the scalar
fields.

The model is defined by functional integral
\eqn\sys{Z[J] = \e{-W[J]} = \sum_{\phi_n \in Z_2} \int \prod_{n \in \Lambda}
d\psi_n d\psibar_n  \e{ -A[\phi, \psi, \psibar] + \sum_n J_n \phi_n}}
with the action
\eqn\act{A[\phi, \psi, \psibar] = - 2 \kappa \sum_{n, \mu} \phi_n
\phi_{n+\hat{\mu}} + \sum_{m, n} \psibar_m (\dsl_{mn} +  y \phi_m
\delta_{mn})\psi_n,}
where $\kappa \in (-\infty, \infty)$ is the hoping parameter and $y \geq 0$
is
the Yukawa coupling;
$\dsl$ is a lattice Dirac operator determining the form of the fermion action
(the system with mirror fermions which we consider in Sect. 4.4 is reduced to
this form).
For all the  formulations under consideration $\dsl$ have the
properties
\eqn\pr{\eqalign{ & \dsl_{mn} = -\dsl_{nm},\cr
& \dsl_{mn} = N^{-D} \sum_{p, \mu} \e{i p (m - n)}
i \, \gamma_{\mu} \, L_{\mu}(p), \cr
& L^*_{\mu}(p) = L_{\mu}(p), \;\;\; L_{\mu}(-p) = - L_{\mu}(p),\cr}}
where $p_{\mu}  = (2 \pi / N) (k_{\mu} - 1/2)$,  $-N/2+1 \leq k_{\mu}\leq
N/2$,
so that $p_{\mu} \in (-\pi/2, \pi/2)$; we use the Hermitean
$\gamma$-matrices:
$[\gamma_{\mu},
\gamma_{\nu}]_{+} = 2 \delta_{\mu \nu}$. In the limit of $N
\rightarrow \infty$ the sum $N^{-D} \sum_p$ defines the  integral
$\int_p \equiv \int^{\pi/2}_{-\pi/2} d^D p / (2 \pi)^D$.

The action \act\ is invariant under $Z_2$ global chiral transformations
\eqn\trans{\phi_n \rightarrow -\phi_n,\; \; \;
\psi_n \rightarrow (-P_L + P_R) \psi_n,\; \; \; \psibar_n \rightarrow
\psibar_n
(-P_R + P_L),}
where $P_{L,R} = (1 \pm \gamma_{D+1})/2$ are chiral projecting operators.

\newsec{The method and approximations}

To analyze the phase diagrams of the system we use the
variational mean field approximation \ref\DZ{J.-M. Drouffe, J.-B. Zuber,
\PREP{102} (1983) 1} (for more detail see ref. \Z) which becomes applicable
to
\sys\ after integrating out the fermions.
Then for free energy of the system $F = W[0]$ it yields the inequality
\eqn\MF{F
\leq F_{MF} = \inf_{h_n} [-\sum_n (u(h_n) - h_n u'(h_n)) - \vev{ 2 \kappa
\sum_{n, \mu} \phi_n  \phi_{n+\hat{\mu}} + \ln \det \,[\dsl +  y \phi]}_h ],}
where $h_n$ is a mean field, and
\eqn\u{\eqalign{& u(h_n) = \ln \sum_{\phi_n \in Z_2} \e{h_n \phi_n} = \ln 2
\cosh h_n, \cr
&\vev{\phi_n} = u'(h_n), \;\;\; \vev{\phi_m\,\phi_n}_h = u'(h_m) u'(h_n) +
\delta_{mn} u''(h_m), \;\;\; etc., \cr
&\ln \det \,[\dsl +  y \phi] \cr
& \;\;\;= \ln \det [\dsl]  -
\sum^{\infty}_{n = 1} {(-1)^{n}\over n} y^n \sum_{i_1,...,i_n} \tr\,
(\dsl^{-1}_{i_1\,i_2}\dsl^{-1}_{i_2\,i_3}...\dsl^{-1}_{i_n\,i_1})\,
\phi_{i_1}\phi_{i_2}...\phi_{i_n} \cr &
\;\;\;= 2^{D/2} N^D \ln y  -
\sum^{\infty}_{n = 1} {(-1)^{n}\over n} {1\over y^n} \sum_{i_1,...,i_n} \tr\,
(\dsl_{i_1\,i_2}\dsl_{i_2\,i_3}...\dsl_{i_n\,i_1})\,
\phi_{i_1}\phi_{i_2}...\phi_{i_n}, \cr}}
where $\tr$ stands for trace over spinorial indices, and in the last
expression the relation $\phi_i^{2^{D/2}} = 1$ has been taken into account.

We consider $F_{MF}$ for two translation invariant
ans\"{a}tze for $h_n$
\eqn\h{\eqalign{& h_n^{FM} = h, \cr & h_n^{AF} = \epsilon_n h,\;\;\;
\epsilon_n =
(-1)^{\sum_{\mu} n_{\mu}}.\cr}}
which, in fact, are the order parameters distinguishing the ferromagnetic
(FM:
$h_n^{FM} \neq 0$, $h_n^{AF} = 0$),  antiferromagnetic (AF: $h_n^{FM} = 0$,
$h_n^{AF} \neq 0$), paramagnetic (PM: both are zero), and ferrimagnetic (FI:
both are nonzero) phases in the system. Then the second order phase
transition lines are determined by equations
\eqn\tl{{\del^2 \over \del h^2}
F_{MF}^{FM,\,AF}\mid_{h = 0} = 0,}
where $F_{MF}^{FM,\,AF}$ is the functional of the right hand side of Eq.
(3.1) on ans\"{a}tze \tl, and therefore, to find the critical lines it is
sufficient to know the functional to terms of order of $h^2$.

Correlations of
$\phi{^,}$s at coinciding arguments (Eq.(3.2)) make the problem unsolvable
exactly, as the contributions of order of $h^2$ to $\vev{\ln \det \,[\dsl +
y
\phi]}_h$ come from terms of any orders of $u''$, as well as from those of
order of
$u'^2$. Therefore, we are forced to use some approximations, and,
particularly,
to use two representations of fermionic determinant (3.2) separately for
``weak"
and ``strong" coupling regimes of $y$, though the exact meaning of this can
only be clear a posteriori.

Our approximation involves summing up all contributions to $\vev{\ln \det
\,[\dsl +  y \phi]}_h$ of the form $u'_{i_1} u'_{i_{n/2+1}} \,
(u'')^{n-2}\,\delta_{i_2 \,i_n} ...
\delta_{i_{n/2} \,i_{n/2+2}}$ (proper ladder diagrams), and
$(u'')^n \, \delta_{i_1\, i_{n/2+1}}  \delta_{i_2\, i_n}$ $ ...
\delta_{i_{n/2}\, i_{n/2+2}}$ (crossed ladder diagrams) (for a graphical
representation see ref.
\Z). Then, using property (2.3) of the Dirac operators we find that  critical
lines in the system in our approximation are determined by the expressions
\eqn\res{\eqalign{
& \kappa^{F(W)}_{cr} = {1 \over 4D} [ 1 - 2^{D/2} ( { y^2 G^W(0) \over 1 +
y^2
G^W (0)} - \int_q  { y^2 G^W (q) \over (1 + y^2 G^W (q))^2})], \cr
&\kappa^{AF(W)}_{cr} = -{1 \over 4D} [ 1 - 2^{D/2} ( { y^2 G^W (\pi)
\over 1 + y^2 G^W (\pi)} - \int_q  { y^2 G^W (q) \over (1 + y^2 G^W
(q))^2})],\cr
&\kappa^{F(S)}_{cr} = {1 \over 4D} [ 1 - 2^{D/2} ( { G^S (0) \over y^2 + G^S
(0)}
- \int_q  { y^2 G^S (q) \over (y^2 + G^S (q))^2})], \cr
&\kappa^{AF(S)}_{cr}
= -{1 \over 4D} [ 1 - 2^{D/2} ( { G^S (\pi) \over y^2 + G^S (\pi)} -
\int_q  { y^2 G^S (q) \over (y^2 + G^S (q))^2})], \cr }}
where
\eqn\g{\eqalign{&G^W (q) =  \int_p {L(p) L(p+q) \over L^2 (p) L^2(p+q)},
\cr &
G^S (q) = \int_p L(p) L(p+q), \cr
& q_{\mu} \in (-\pi, \pi], \;\;\; \int_q \equiv \int^{\pi/2}_{-\pi/2} {d^D q
\over {(2
\pi)^D}}. \cr}}
The contributions to (3.5) which are proportional to $G(0)$ or
$G(\pi)$ come from the proper ladder diagrams, while the integral terms from
the cross ladder ones.

As $G(0) > 0$ and $G(\pi) < 0$,  integral terms in (3.7) for $D = 4$ are
logarithmically divergent at $y^2 = -1/G^W(\pi)$ or $-G^S(\pi)$. Therefore,
even though in weak coupling regime they are of  rder of $D^{-1}$ compared
with the first ones, they cannot be neglected at least in the vicinity of
these
points. Thereby these terms determine domains of the ``weak" and ``strong"
coupling regimes also for $\kappa^F_{cr}$. They are domains of analyticity of
functions
$\kappa^W_{cr}(y)$ and $\kappa^S_{cr}(y)$, that is $y^2 < -1/G^W (\pi)$ and
$y^2 > -G^S (\pi)$, respectively, coinciding for $\kappa^F_{cr}$ and
$\kappa^{AF}_{cr}$. The integral terms are always negative and
therefore increase the contributions of the proper ladder terms for
$\kappa^F_{cr}$ and decrease them for
$\kappa^{AF}_{cr}$.  In what follows we shall approximate these terms by
functions $c^W y^2 \ln [1 + y^2 G^W(\pi)]$ and $c^S \ln [1 + G^S(\pi) / y^2]$
with some positive constants $c^W$ and $c^S$.

Contributions of other diagrams to $\vev{\ln \det \,[\dsl +  y \phi]}_h$ come
into play in higher orders in $y^{\pm 2}$, at least from the order of $y^{\pm
6}$. So, even though in strong coupling regime they can give contributions to
$\kappa_{cr}$ of the same order in $1/D$ as the ladder ones, the assumption
that their contributions are suppressed and non-singular looks plausible.

\newsec{}

In this section we apply formulae (3.5) for $D = 4$ to the system with
different chirally invariant formulations of fermions on a lattice.

\subsec{Naive fermion action}

In this case operator $\dsl$ is local
\eqn\naive{\eqalign{&\dsl_{mn} = \sum_{\mu} \gamma_{\mu} \,{1\over 2}
(\delta_{m + \hat{\mu} \; n}  - \delta_{m - \hat{\mu} \; n}),\cr
 &L_{\mu}(p) = \sin p_{\mu},\cr}}
and therefore produces species doubling.

Due to well known
symmetry of the model under transformations:
$(\psi, \psibar)_n$ $\rightarrow \exp(i \epsilon_n \pi /4) (\psi,
\psibar)_n$,
$\phi_n \rightarrow \epsilon_n \phi_n$, $\kappa \rightarrow -\kappa$, $y
\rightarrow -i y$, one has $G(\pi) = -G(0)$. Numerically $G^W(0) \simeq
0.62$,
$G^S(0) = 2$, and therefore the domains of weak and strong
coupling are $y\;
\roughly{<}\; 1.27$ and $y > 2^{1/2}$, respectively. The phase diagram is
shown
in Fig. 1. The lines
$\kappa^F_{cr}(y)$ and $\kappa^{AF}_{cr}(y)$ do not intersect each other, so
no FI phase appears (in this point we agree with previous results
\ref\STs{M. A. Stephanov, M. M. Tsypin, \PL{236} (1990) 344}, \ref\EK{T.
Ebihara, K.-I. Kondo, Nucl. Phys. B (Proc. Suppl.) 26 (1992) 519}). The lines
form two disconnected domains with PM phase(s), as well as with AF phase(s).

\subsec{SLAC-like fermion action}

This action is defined by operator $\dsl$ of the form
\eqn\SLAC{\eqalign{&\dsl_{mn} = \sum_{\mu} \gamma_{\mu} \, \sum_{l >
0}(-1)^{l+1} {1\over N} ({1\over{\sin \pi(l+{1\over2})/N}} \cr
& \;\;\;\;\;\;\;\;\; + {1\over{\sin \pi(l-{1\over2})/N}}) \,
(\delta_{m + l\hat{\mu} \; n}  - \delta_{m -
l\hat{\mu}
\; n}),\cr  &L_{\mu}(p) = 2 \sin {1\over2}p_{\mu}.\cr}}
It represents an action with a moderate non-locality as $\dsl_{mn}$ drops
like $|m - n|^{-1}$ \ref\SLAC{S. D. Drell, M. Weinstein, S.
Yankielovitcz,
\PRD{14} (1976) 487}.

In this case we have $G^W(0) \simeq 0.15$, $G^W(\pi)
\simeq -0.095$; $G^S(0) = 8$, $G^S(\pi) = -16/\pi$. The new point here is
t`at
the domains of weak and strong coupling are overlapped: $y^W\;
\roughly{<}\; 3.24$,
$y^S > 4 \pi^{-1/2} \simeq 2.26$, so, we can continue lines
$\kappa^{F(W)}_{cr}(y)$ and $\kappa^{F(S)}_{cr}(y)$ until they intersect each
other. Then phase diagram looks like in Fig. 2. No
FI phase appears, but in contrast to the local case both domains with PM and
AF
phases are connected.

\subsec{Weyl fermion action}

This action is defined by the finite dimensional approximation of functional
integrals for Weyl quantization \ref\Ze{S. V. Zenkin, \MPL{6}
(1991) 151}. In this case we have
\eqn\Weyl{\eqalign{&\dsl_{mn} = \sum_{\mu} \gamma_{\mu}\, \sum_{l >
0}(-1)^{l+1}\, 2\, (\delta_{m + l\hat{\mu} \; n}  - \delta_{m - l\hat{\mu} \;
n}),\cr
&L_{\mu}(p) = 2 \tan {1\over2}p_{\mu}.\cr}}
This is a maximally non-local action in the sense that $\dsl_{mn}$
is merely an alternating sum over the whole lattice and does not drop at all
with increasing $|m-n|$. But a remarkable fact is that this action can be
transform to a locam form if we introduce variables $\psi^d$ defined on the
centres of D-cells of the lattice, i.e. on the sites of the dual lattice,
leaving fields $\psibar$ being defined on sites of the original one. Then
change of variables, which in momentum space looks like
$\psi_p = F(p) \psi^d_p$, with $F(p) = \prod_{\mu} \cos {1\over2}
p_{\mu}$, leads to a local action with
$L_{\mu}(p) = 2 \sin {1\over2} p_{\mu} \prod_{\nu \neq \mu} \cos
{1\over2} p_{\nu}$. Although now kinetic part of the action produces
$2^{D-1}$-fold species doubling, the system of course is not changed:
contributions of the additional species to the partition function are
canceled
by Jacobian coming from the change of variables, while their coupling to the
Higgs field is suppressed by the factor $F(p)$ (in this point this looks
similar
to the Zaragoza proposal \ref\Aea{J. L. Alonso, J. L. Cort\'{e}s, F. Lesmes,
Ph. Boucaud, E. Rivas, Nucl. Phys. B (Proc. Suppl.) 29B,C (1992) 171}).

The more non-locality of the action, the less $|G^W|$ and the grater $|G^S|$.
In
this extremal case $G^W(0) \simeq 0.045$, $G^W(\pi)\simeq -0.0074$, at $N
\rightarrow \infty$ we have $G^S(0) \rightarrow \infty$,
$G^S(\pi) = -16$. Like in the case of the SLAC-like action domains of weak
and
strong coupling regimes are overlapped:
$y\;\roughly{<}\; 11.6$, $y > 4$,
but the phase diagram shown in Fig. 3 is qualitatively different. FI phase
appears which is spreaded to any $y <
\infty$, all four phases being connected.

\subsec{Mirror fermion action}

We consider the simplest variant of the mirror fermion action \ref\Mm{I.
Montvay, \PL{199B} (1987) 89; Nucl. Phys. B (Proc. Suppl.) 29B,C (1992) 159}
with zero mixing parameter between fermion field $\psi$ and its mirror
counterpart
$\chi$ and with only $\chi$ coupled to the Higgs field
\eqn\mf{\eqalign{A = \sum_{n, \mu}\, [ \, & \psibar_n \gamma_{\mu}
{1\over 2} (\psi_{n +
\hat{\mu}} - \psi_{n - \hat{\mu}}) - {1\over 2} \psibar_n (\chi_{n +
\hat{\mu}} + \chi_{n - \hat{\mu}} - 2 \chi_n) \cr
+ \, & \overline{\chi}_n \gamma_{\mu} {1\over 2} (\chi_{n +
\hat{\mu}} - \chi_{n - \hat{\mu}}) - {1\over 2} \overline{\chi}_n (\psi_{n +
\hat{\mu}} + \psi_{n - \hat{\mu}} - 2 \psi_n)] \cr
 + \sum_{n} \, & y \, \overline{\chi}_n \phi_n \chi_n. \cr}}
The action has the mirror symmetry
\eqn\mfs{\eqalign{ & \psi_n \rightarrow (-P_L + P_R) \psi_n,\; \; \;
\psibar_n
\rightarrow \psibar_n (-P_R + P_L), \cr
& \chi_n \rightarrow (-P_R + P_L) \chi_n,\; \; \; \overline{\chi}_n
\rightarrow
\overline{\chi}_n (-P_L + P_R), \cr
& \phi_n \rightarrow -\phi_n.}}
We can apply our formulae to this system after integrating out $\psi$. Then,
as $\psi$ does not couple to $\phi$ and therefore its determinant is an
irrelevant constant, we come to effective non-local action in terms of fields
$\chi$ and $\phi$ of the form of (2.2) with
\eqn\ml{L_{\mu}(p) = \sin p_{\mu} \, [ 1 + {(\sum_{\nu} ( 1
- \cos p_{\nu} ) )^2 \over \sum_{\nu} \sin^2 p_{\nu}}].}

This non-locality is of a new type compared with two preceding cases as
$L_{\mu}(p)$ now involves also all $p_{\nu}$ with $\nu \neq \mu$. Now we have
$G^W(0) \simeq 0.026$, $G^W(\pi)\simeq -0.0073$; $G^S(0) \simeq 350$,
$G^S(\pi) \simeq -160$. So this case incorporates features of both local and
non-local actions: $|G^W|$ are small and $|G^S|$ are big, but the
domains of weak and strong coupling regimes $y\; \roughly{<}\; 11.7$, $y \;
\roughly{>}\; 12.6$ are not overlapped. The phase diagram is shown in Fig. 4,
it has the same features as that for naive fermion action.

\newsec{Conclusion}

These results demonstrate that the phase structure of the
$Z_2$ Higgs-Yukawa system is very sensitive to the form of lattice fermion
action used for its exact definition, even though the Higgs field couples to
fermions in the same way. The interesting question is whether this is so for
continues groups \ref\TZ{S. Tominaga, S. V. Zenkin, in progress}.

Another point is that we only found the second order phase transition lines,
and
did not investigate the questions concerning the first order phase
transitions
and particle contents of the phases. Meanwhile the latter issue may have
interesting consequences for definition of chiral gauge theories on a
lattice.
It was found (for a review and further references see
\ref\Sh{J. Shigemitsu, Nucl. Phys. B (Proc. Suppl.) 20 (1991) 515}, and also
\STs) that in the Higgs-Yukawa models with naive fermions the latter
acquire masses of order of inverse lattice spacing at the right wing of PM
phase, i.e. in the strong Yukawa coupling regime. The masses do not go to
zero even at
PM-FM boundary where continuum limit of the system can be defined, so the
fermions decouple
from the spectrum. The mirror fermion system (4.4) can be considered as a toy
model of a
world with light fermions and heavy mirror fermions. If along the right wing
of
PM-FM line of Fig. 4 the mirror fermions also acquire masses of order of
$O(1)$,
while the fermions do not, there appears a possibility for nonperturbative
definition of chiral gauge theories, which to our best knowledge still was
not
explored.

\bigbreak\bigskip
\centerline{{\bf Acknowledgments}}
\vskip 8pt

This work is supported by JSPS. I am grateful to H. Yoneyama for useful
discussions and J. L. Alonso for clearing up some points of Zaragoza
proposal.
It is also a pleasure to thank all members of the Elementary Particle Theory
Group of the Department of Physics of Kyushu University for their warm
hospitality.

\listrefs

\vfil\eject
\epsfxsize=\hsbody
\multiply \epsfxsize by 9
\divide \epsfxsize by 10
\centerline{\epsfbox{Fig1.eps}}
\vskip 0.4cm
Fig. 1. Phase diagram of the system with naive fermion action (4.1); solid
lines are FM-PM phase transition lines, gray ones are PM-AF phase
transition lines.
\vskip 0.4cm
\epsfxsize=\hsbody
\multiply \epsfxsize by 9
\divide \epsfxsize by 10
\centerline{\epsfbox{Fig2.eps}}
\vskip 0.4cm
Fig. 2.  Phase diagram of the system with SLAC-like fermion action (4.2).

\vfil\eject
\epsfxsize=\hsbody
\multiply \epsfxsize by 9
\divide \epsfxsize by 10
\centerline{\epsfbox{Fig3.eps}}
\vskip 0.4cm
Fig. 3. Phase diagram of the system with the Weyl fermion
action (4.3).

\vskip 1.3cm
\epsfxsize=\hsbody
\multiply \epsfxsize by 9
\divide \epsfxsize by 10
\centerline{\epsfbox{Fig4.eps}}
\vskip 0.4cm
Fig. 4. Phase diagram of the system with mirror
fermion action (4.4).
\bye